\documentclass[aps,pre,bibtex,twocolumn,notitlepage,superscriptaddress]{revtex4-2}

\usepackage{epsfig}
\usepackage{graphicx}
\usepackage{amsmath}
\usepackage{amssymb}
\usepackage{color}

\usepackage{hyperref}
\hypersetup{colorlinks=true,linkcolor=blue,citecolor=blue}
\usepackage{lipsum}  
\usepackage{physics}

\newcommand{\ve}[1][K]{\mathbf{#1}}

\usepackage{ulem}

\begin{document}

\title{Optimization of  multisite reactions in complex compartmentalized media}

\author{T. V. Mendes}
\affiliation{Laboratoire Ondes et Mati\`ere d'Aquitaine, CNRS/University of Bordeaux, F-33400 Talence, France}
\affiliation{Centre de physique théorique, CNRS / Aix-Marseille university, F-13288 Marseille, France}

\author{T. Gu\'erin}
\affiliation{Laboratoire Ondes et Mati\`ere d'Aquitaine, CNRS/University of Bordeaux, F-33400 Talence, France}

\begin{abstract}
In complex media, transport and geometric properties deeply influence the kinetics of random encounters between reactants. Here, we consider the situation where a random walker, moving in a regularly diffusing medium, has to 
reach and activate a target located inside a compartment characterized by fractal (obstructed) sub-diffusion. We focus on dual-site reactions, which end when two activation events occur  within a given time window. Each activation event happens with a finite probability whenever the random walker visits the target. For weakly reactive targets, we demonstrate that the reaction time can be minimized for an optimal compartment size and can even be accelerated when compared to the same system without compartment. Our analytical predictions are validated through simulations of a random walker on a cubic lattice, where some sites inside the compartment are obstructed at the critical percolation threshold. Our theory illustrates the fact that adding a crowded compartment around a target, even if it slows down the motion in its vicinity, can accelerate the kinetics of complex reactions, especially for weakly reactive targets.
\end{abstract}

\maketitle

\section{Introduction}
The kinetics of first passage, which quantifies the time a random walker needs to reach a target, has been intensively studied in the last decades with applications in several areas of science, ranging from biophysics to the kinetics of transport influenced reactions and search processes \cite{Redner:2001a,metzler2014first,grebenkov2023diffusion,benAvraham2000,benichou2011intermittent,loverdo2008enhanced,weiss2014crowding,Condamin2007}. The first passage time is a useful quantity when the concentration of searchers (or targets) is so low that standard chemical rate equations are no longer reliable. Previous studies have shown a non-trivial interplay between the first passage time distribution and the properties of transport and geometry \cite{Condamin2007,Benichou2010,godec2016universal,grebenkov2018towards,grebenkov2019full,grebenkov2018strong,Schuss2007}.

So far, up to the exception of diffusive transport \cite{mangeat2021narrow,vaccario2015first,dosSantos2022efficiency}, theoretical approaches of first passage in confinement have focused on transport in  {homogeneous} media, leaving aside the case of media presenting sub-compartments  with different transport properties. 
However, compartmentalized media are ubiquitous in nature,  as evidenced by membrane-less organelles (MLOs) 
in various biological (and synthetic) systems,
which may be involved (among other functions) in the regulation of (bio)chemical reactions \cite{andre2020liquid,nakashima2019biomolecular,oFlynn2021role}.
To the difference of membrane-bounded compartments, where the concentration of reactants can be regulated by membrane pumps and channels, in MLOs reactants can 
diffuse across the interface between the compartment and the surrounding medium. These organelles are typically dense, crowded, much more viscous than the surrounding medium \cite{elbaum2015disordered,wang2021surface}, and display complex viscoelastic \cite{michieletto2022rheology,alshareedah2021programmable} and transport properties~\cite{watanabe2024diffusion,shakya2018non,shen2023biological,munoz2022stochastic}. 
 
The fact that transport inside these compartments is slower  than outside them suggests that they act as kinetic barriers that must be overcome to enable reactions, for example, by over-concentrating the reactants inside them. 
One  could however wonder whether the different nature of the  transport properties inside   compartments could accelerate reaction kinetics.   
Indeed, it is well established that target search kinetics can be optimized by combining distinct transport phases, as exemplified by the search for a target sequence on DNA \cite{coppey2004kinetics,von1989facilitated,Berg1985}, intermittent search strategies~\cite{benichou2011intermittent} or target search acceleration by combining surface-mediated and bulk diffusion \cite{calandre2014accelerating,benichou2010optimal}. 

Here, we ask the question as to whether a crowded compartment, displaying fractal-subdiffusion properties, could optimize the kinetics of multisite reactions. 
Multisite reactions are ubiquitous in biological media~\cite{van2007allosteric,nash2001multisite,miller2005stability,crabtree2002nfat,takahashi2010spatio,hellmann2012enhancing,aoki2011processive,gopich2013diffusion} and in particular in phosphorylation reactions. 
In a multisite reaction, a target displays several binding sites which need to be simultaneously  active 
to trigger a reaction.  An activation event is  obtained with some probability when a random walker hits the target (and may deposit a molecule on the binding site, for example a phosphate for phosphorylation reactions).  
Typically, binding sites remain active only transiently (bounded molecules can be degraded or released), so that multisite reactions  can only occur when all the required activations happen  within a given time window. Intriguingly, slower transport properties can increase multisite reaction efficiency, either for diffusive~\cite{gopich2013diffusion,takahashi2010spatio} or complex~\cite{aoki2011processive,hellmann2012enhancing} transport; however these theories have not considered confined space and in particular compartmentalized media. This is the subject of the present work.

Here, we  investigate whether dual-site reactions can be accelerated in  a simple model of complex compartmentalized media. Considering a target inside a compartment displaying fractal (obstructed) sub-diffusion~\cite{weiss2014crowding}, it-self surrounded by a diffusive confined medium, we calculate the mean time a random walker needs to  activate the target twice within a given time window. We show that, for weakly reactive targets, the reaction time can be minimized for an optimal compartment size. 
This result is demonstrated   for  a cubic lattice for which some sites are obstructed inside the compartment at the percolation threshold. We find that crowding can even lead to reaction times that are much smaller than those for a lattice without any obstacles (for which transport is much faster). This optimization effect comes from the higher probability of subsequent hits to the target due to compact search, which compensates and even overcomes the crowding-induced slow-down of transport inside the compartment. 
Finally, we show that this optimization effect is robust under a certain number of generalizations, such as with mobile targets, several random walkers, or multi-site reactions with more than two sites.
  
\section{Model} 

We consider a random walker moving on a  discrete network of $N$ sites of positions $\ve[r]_i$, with $i\in \{1;N\}$. 
The waiting time on a site $i$ is    exponentially distributed with average $\nu_i^{-1}$. At each jump, the next site is chosen randomly among the neighbors of the current site. We define $P(\ve[r]_i,t\vert\ve[r]_j)$, the probability  that the random walker is at  $\ve[r]_i$ at time $t$, given that it starts at $\ve[r]_j$ at time $0$. We consider an ``edge centric dynamics'' \cite{masuda2017random}, for which $\nu_i$ is proportional to the number of neighboring sites, ensuring that the stationary probability is uniform, 
$\lim_{t\to\infty} P(\ve[r]_i,t\vert\ve[r]_j)=1/N$. 
Assuming that the network is embedded in a three-dimensional  space, we call $ \vert \ve[r]_i-\ve[r]_j\vert $ the Euclidian distance between two sites (see \textcolor{blue}{Appendix \ref{Dist}} for a discussion on the choice of this distance). We assume that this network displays a compartment-like structure. When the distance $r$ to a reference site (the central site, taken at the origin $\ve[r]_T=\ve[0]$) is  smaller than a fixed value $R$, the random walk displays the properties of fractal subdiffusion, with a walk-dimension $d_{\text{w}}$ defined such that $\langle r^2(t)\rangle\sim K t^{2/d_{\text{w}}}$, and a spatial dimension $d_{\text{f}} $ so that the number of sites at distance $r$ from a reference site is $n(r)\sim r^{d_{\text{f}}}$.  This class of models has often been used as a first step to describe anomalous transport in crowded media in synthetic \cite{zunke2022first} or biological environments \cite{saxton2008biological,Benichou:2011,hofling2013anomalous,weiss2014crowding,szymanski2009elucidating}, some of them (cytoplasm, chromatin, bones...) displaying fractal properties \cite{bancaud2009molecular,pothuaud2000fractal,aon1994fractal,szymanski2009elucidating}.
Distances $r<R$ define the compartment, while for $r>R$ we assume a classical three-dimensional diffusive random walk, for which the walk dimension is equal to two. We assume that compartments are much smaller than the confining embedding medium, and that the transport inside them is compact, with $d_{\text{f}}<d_{\text{w}}$. Note that the assumption that all sites have the same stationary occupation probability, including in the compartment, means that there is no energy gain for the random walker to be in the compartment.

We consider the central site as an imperfectly reactive target. Each time the random walker visits the target, the visit 
is considered as ``successful'' with probability $p$, in which case the target switches into an ``active'' state in which it remains during a time $\Delta$. We define the activation time $\tau_{\ve[r]}$ as the time  of 
first successful passage to the target when the initial position is $\ve[r]$, $\tau_1$ is the same quantity when the starting position is a neighbor of the central site.  We call $\tau_{\ve[r]}^*$   the first passage time (FPT). 
We assume that a second successful visit to an active target triggers a  reaction. The reaction time $T$   is  defined as the time to obtain two successful target hits at times separated by less than $\Delta$, for the first time. This is a minimal model for dual-site reactions, in which a target presents two binding or reactive sites that have to be simultaneously active to trigger a reaction, $\Delta$ representing the  time after which a single site is inactivated. 

\section{Mean first passage and activation times}
  
Since the transport inside the compartment is compact ($d_w>d_f$), we expect that the mean first passage time depends on the initial distance $r$ when $r\ll R$. 
To calculate it, we adapt the arguments of Refs~\cite{Condamin2007,reuveni2010vibrational}. 
We start with the exact expression~\cite{noh2004random,Condamin2007}
\begin{equation}
\langle \tau_{\ve[r]}^* \rangle   = N \int_0^\infty dt\  [P(\ve[r]_T,t\vert \ve[r]_T)-P(\ve[r]_T,t\vert \ve[r])].  \label{ExactMFPT}
\end{equation}
Following Ref.~\cite{Condamin2007}, we consider the large volume limit $N\to\infty$, meaning that the confining region is much larger than the compartment. In this limit,  the propagators $P$ in the above integral can be evaluated in infinite space. If $r=\vert  \ve[r]_T-\ve[r] \vert \ll R$, the  integral (\ref{ExactMFPT}) is dominated by the values $t$ that are small enough so that the boundary of the compartment is not reached, for which the propagator satisfies the standard scaling \cite{benAvraham2000}
\begin{equation}
P(\ve[r]_T, t\vert \ve[r]) \simeq \frac{1}{t^{d_{\text{f}}/d_{\text{w}}}}\Pi\left(\frac{r}{t^{1/d_{\text{w}}}}\right), \label{ScalingFractal}
\end{equation} 
where the positive function $\Pi$ decays for large arguments. 
Evaluating  the integral (\ref{ExactMFPT}) with the form (\ref{ScalingFractal}) for the propagators leads to 
\begin{equation}
 \langle \tau_{\ve[r]}^* \rangle \underset{ r\ll R}{\simeq} A\  N \ r^{d_{\text{w}}-d_{\text{f}}} \label{ScalingMFPTInitialDistance},
\end{equation}
where the prefactor  $A=\int_0^\infty du\ u^{-d_{\text{f}}/d_{\text{w}}} [\Pi(0)-\Pi(1/u^{1/d_{\text{w}}})]$ does not depend on the geometry of the system. This expression extends the scalings of Refs.~\cite{Condamin2007,reuveni2010vibrational}  to compartmentalized media. 

The above formula indicates that, the mean time to find the target starting from $r=R$ scales as $ \langle \tau_{R}^* \rangle\sim N R^{d_w-d_f}$. Next, the time to reach the compartment starting from $r>R$ scales as \cite{Benichou2008} 
$C N \left(\frac{1}{R}-\frac{1}{r}\right)$, with $C$ a constant \footnote{For a cubic lattice of lattice spacing $a$, $C=1/(4\pi D)$, with $D=\nu a^2/6$.}. This time saturates at large $r\gg R$ and  is much smaller for large $R$ than the time $\langle \tau_R^*\rangle$ to find the target starting from $R$. Since we assume that there is no energy barrier at the compartment boundary, $\langle \tau_{\ve[r]}^*\rangle$ is continuous at $r=R$. Therefore, if the random walker starts outside the compartment, the mean first passage time reads 
\begin{align}
\langle \tau_\infty^*\rangle=\alpha\  N  R^{d_{\text{w}}-d_{\text{f}}},  \label{MFPTINF}
\end{align}
with a prefactor $\alpha$ that does not depend on $R,N$. Next, when the starting position is a neighbor of the target site, 
Kac's theorem \cite{aldousFill2014} leads to
\begin{equation}
\langle\tau_1^*\rangle = N/\nu,  \label{MFPT1}
\end{equation}
with $\nu^{-1}=\nu_T^{-1}$. 
One can use these expressions for the mean first passage times to obtain mean activation times (when $p<1$) by writing  $\langle \tau_{\ve[r]} \rangle = \langle \tau_{\ve[r]}^*\rangle+\langle (n-1) \tau_1^*\rangle$ where $n$ is the number of passages required to obtain an activation event. The probability of an activation at the $n^{\text{th}}$ passage is $p(1-p)^{n-1}$, so that $\langle n-1\rangle=(1-p)/p$. Therefore, 
\begin{equation}
\langle \tau_\infty \rangle = \alpha \ N \ R^{d_w-d_f} +\frac{(1-p )N}{p\nu}, \hspace{0.5cm}\langle \tau_1 \rangle =  \frac{N }{p\nu}.\label{MFPTImp}
\end{equation}

\section{Multisite reactions: optimization} 

We now investigate the properties of the reaction time $T$, previously defined as the time to obtain two  activations separated by less than $\Delta$, when the initial position is outside the compartment. If $\Delta\gg\langle\tau_\infty\rangle$, the target is always still active when the second activation occurs, so that $\langle T\rangle\simeq \langle\tau_\infty\rangle+\langle\tau_1\rangle$, and acceleration of the multisite reaction is impossible. We thus focus on the opposite case $\Delta\ll \langle\tau_\infty\rangle$ (more general formulas can be found in Appendix \ref{Deriv_Eqs}). Let $Q$ be the probability that, after an activation, the reaction ends before the random walker escapes far from the compartment. If the volume is large enough, every trajectory that remains close to the compartment lasts a time that is negligible compared to $\langle \tau_\infty\rangle$, so that $T=\tau_\infty$ with probability $Q$. In turn, with probability $1-Q$, the random walker escapes without completing the reaction,  and the search for an event formed by two successful and consecutive passages has to start over entirely,  so that $T=\tau_\infty+T'$, with $T'$ an independent  copy of $T$. 
This leads to $\langle T\rangle=\langle\tau_\infty\rangle Q+ (\langle\tau_\infty\rangle+\langle T\rangle) (1-Q)$, so that
\begin{align}
\langle T\rangle=\frac{\langle\tau_\infty\rangle }{Q}.\label{TvsQ}
\end{align}

We first consider the case where $\Delta$ is larger than the typical time to go out of the compartment $\tau_c\equiv (R^2/K)^{d_w/2}$, so that after $\Delta$  the searcher is typically in the external region.  In this case, $Q$ is the weight of the direct trajectories, which activate the target without going to large distances. Since the random walk at long times is diffusive in three dimensions, it belongs to the class of non-compact walks for which this probability can be expressed in terms of a ratio of activation times~\cite{godec2016universal,Benichou2010,grebenkov2018strong,guerin2021universal}: 
\begin{align}
1-Q=\langle \tau_1\rangle/\langle \tau_\infty\rangle \label{ValueQ}.
\end{align}
 
We focus on the limit $p\ll1$, which is relevant if the target represents a sphere with a small number of reactive patches,  or if reaching the active state necessitates overcoming an energy barrier \cite{grebenkov2023diffusion,grebenkov2023diffusion}, or the binding of additional molecules \cite{nguyen2024competition}, 
so that each target visit is unlikely to trigger an activation event. In this limit, using  (\ref{MFPTImp}), (\ref{TvsQ})  and (\ref{ValueQ}), we finally obtain
\begin{equation}
\langle T \rangle   \simeq  \frac{\langle\tau_\infty\rangle^2}{\langle\tau_\infty^*\rangle-\langle\tau_1^*\rangle} = N \frac{\left(\alpha R^{d_{\text{w}}-d_{\text{f}}}+\frac{1}{p\nu}\right)^2}{ \alpha R^{d_{\text{w}}-d_{\text{f}}}}, \label{MeanTComp}
\end{equation}
which is independent of $\Delta$. This formula holds when $\tau_c \ll \Delta\ll  \langle\tau_\infty\rangle $, which represents a wide range of $\Delta$ for large $R,N$ and small $p$. 
$\langle T\rangle$ admits a minimum $T_m$ at a value $R=R_m$, with 
\begin{equation}
R_m=\left(\frac{1}{p\nu\alpha}\right)^{\frac{1}{d_{\text{w}}-d_{\text{f}}}}, \ \ T_m=\langle T \rangle_{R=R_m} = \frac{4 N }{p\nu}\label{OptT}.
\end{equation}
Remarkably, $R_m$ diverges when $p$ vanishes, meaning that for weakly reactive targets
multisite reactions can be accelerated for  large compartment radii, although this also enlarges the region with slow transport  around the target. The optimization comes from the fact that, on the one hand, large values of $R$ are obviously not beneficial as they would slow down the motion, while, on the other hand, small values of $R$ increase the probability of escaping to large distances after the first activation without hitting again the target, thus leading to resetting the search too often. The optimal value  $R_m$ finds the balance between these two effects. 

For smaller values of $\Delta$, for which the random walker is still inside the compartment at a time $\Delta$ after an activation event, when $p\ll 1$, we show in  Appendix \ref{DerivSmallDelta} that 
\begin{align}
\langle T\rangle\simeq  \frac{ b N}{p^2\Delta^{1-\frac{d_\text{f}}{d_\text{w}}}}\label{TSmallerDelta}
\end{align}
where $b$ does not depend on $N,R,p$. The above formula, valid for $\nu^{-1}\ll\Delta\ll\tau_c$, means that $\langle T\rangle $ decreases with $\Delta$ until reaching the plateau described by Eq.~(\ref{MeanTComp}).
 
\begin{figure*}
\includegraphics[width=\linewidth]{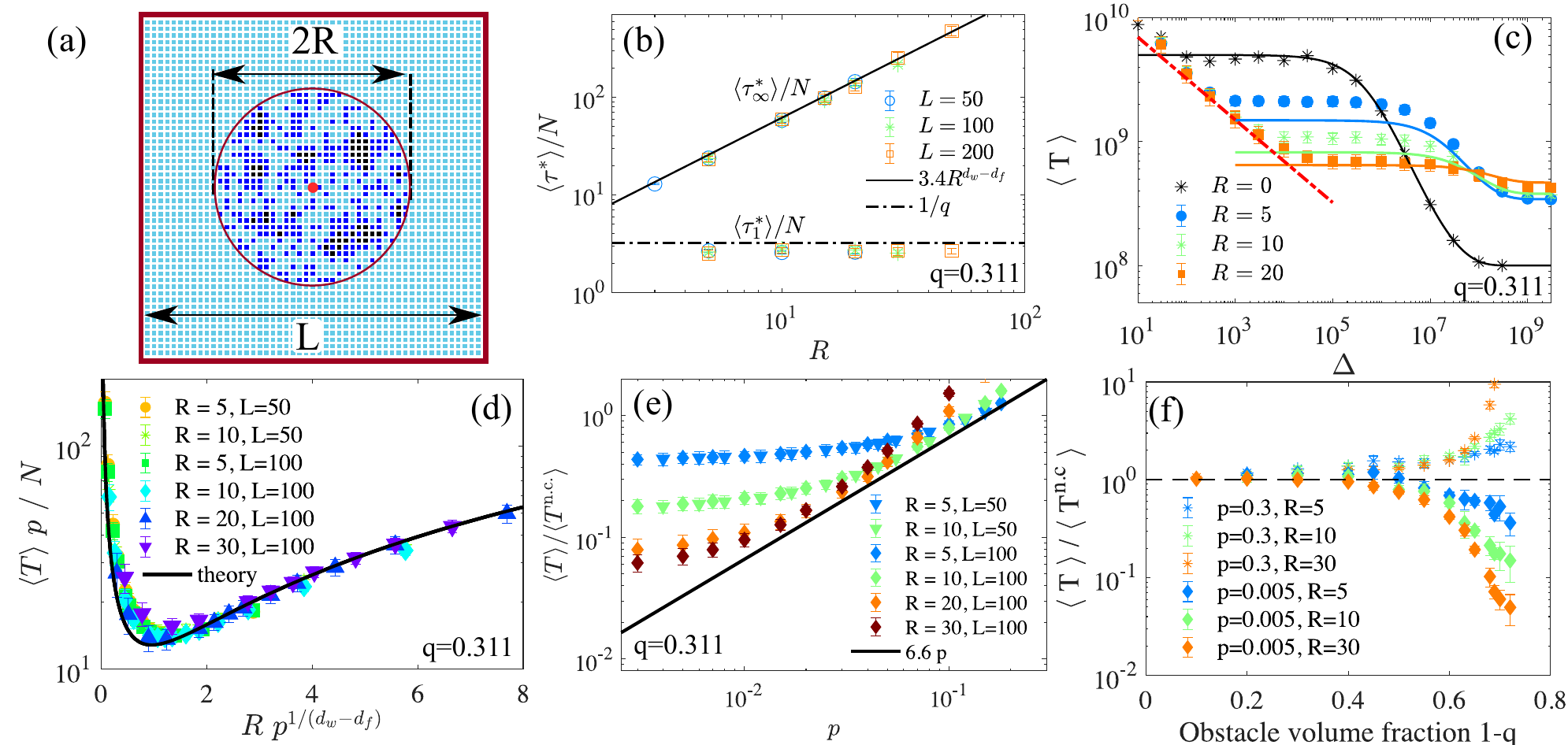}
\caption{(a) Sketch of the system investigated in our simulations, consisting of a crowded sphere of radius $R$ embedded in a regular lattice of side $L$ (the surrounding medium). Light blue : free sites that are connected to the surrounding medium. Deep blue: obstructed sites. Black: free sites that are not connected to the surrounding medium. Red:  target site. Here, the system   is in two-dimensions, with $q=0.59$, but our simulations are in three dimensions.  
(b) $\langle \tau_\infty^*\rangle$ and $\langle \tau_1^*\rangle$ as a function of the compartment size. Symbols: simulations, with $r=23$ for $L=50$, $r=44$ for $L=100$, $r=88$ for $L=200$. 
Continuous black line: scaling (\ref{MFPTINF}), with $\alpha\simeq 3.4$ (obtained from a fit). Dash-dotted black line: $\langle \tau_1^*\rangle \simeq N/\nu$ with $\nu\simeq q$.  
(c) Average reaction time $\langle T\rangle$ as a function of $\Delta$. Symbols: simulation results for $L=100$, $p=0.02$ and $r=44$. Lines with corresponding colors are  the theoretical predictions (\ref{TNotSmallDelta}). 
Black line: theoretical value for $R=0$ obtained from Eqs.~(\ref{MFPTNC}) and (\ref{TNotSmallDelta}). 
The dash-dotted line represents (\ref{TSmallerDelta}) with $b=6$.
(d) $\langle T\rangle$ versus $R$ in rescaled variables. Symbols: simulations with $\Delta=10^4$ for $L=50$ and $\Delta=10^5$ for $L=100$. Black line: theoretical prediction  (\ref{MeanTComp}). 
(e) Acceleration factor $\langle T\rangle/\langle T^{\text{n.c.}}\rangle$ as a function of $p$  for the same data as in (d). Black line: prediction  for the minimal acceleration factor from Eqs.~(\ref{MeanTComp}) and (\ref{DiffRes}).
In (b)-(e), $q=0.311$. 
(f) Acceleration factor versus obstacle density $1-q$ for $\Delta=10^5$ and $L=100$.
All quantities are averaged over the disorder, see Appendix \ref{Simulations} for estimates of confidence intervals. 
The unit of length is the lattice spacing, the unit of time is the residence time on a site without obstructed neighbors. 
}
\label{Fig2RW}
\end{figure*}

\section{Simulations}   

In our simulations, we start with a  regular cubic lattice of size $L\times L \times L$; the lattice spacing is taken as the unit of length. In a spherical region of radius $R$, each site is either kept with probability $q$ or removed (obstructed) with probability $1-q$,  see  Fig.~\ref{Fig2RW}(a). 
At each time step (taken as the unit of time), a random walker attempts to move to one of the neighboring sites (obstructed or not). If the site is not obstructed the move is accepted, and rejected otherwise.  
We consider a target site at the center of the compartment (if it is not connected to the external medium the configuration is  rejected).  For a searcher starting far from the compartment, we measure reaction times  averaged over many realization of the disorder.

When $q=q_c\simeq 0.311$, in an infinite medium, connected sites form clusters with fractal properties: $d_{\text{f}}\simeq 2.523$ \cite{jan1998random,ballesteros1999scaling,deng2005monte,benAvraham2000} and $d_{\text{w}}\simeq3.78$ 
\cite{argyrakis1984random,blavatska2008walking,ben1982diffusion,havlin1983diffusion}. In Appendix \ref{Simulations}, we check that these values also describe the spatial and transport properties inside the compartment for the sites  connected to the external medium.  
The scaling forms (\ref{MFPTINF}) and ~(\ref{MFPT1}), for the mean first passage time starting far from the target and at distance one from it, are then validated in Fig.~\ref{Fig2RW}(b). 

 Fig.~\ref{Fig2RW}(c) shows $\langle T\rangle$ as a function of $\Delta$, where one clearly sees the presence of a plateau, as expected [see Eq.~(\ref{MeanTComp})]. For smaller values of $\Delta$, the results are compatible with the scaling  (\ref{TSmallerDelta}). The curves $\langle T\rangle $ versus $R$ are shown in Fig.~\ref{Fig2RW}(d) and confirm the existence of an optimal compartment radius. Data for different $L,R$ and $p$ can be collapsed on a single   curve using rescaled variables, in agreement with the prediction (\ref{MeanTComp}). Note that the agreement between theory and simulations is not perfect, this may come from the fact that the distribution of $\tau_\infty$, due to the disorder, is slightly different from an exponential distribution.

We also consider the case without compartments (``No Compartment'', n.c.), defined as the same system when all sites are free.  
In this case, the mean activation  times can be obtained exactly in the large $N$ limit \cite{Condamin2005,guerin2021universal}:
\begin{equation}
 \langle \tau_\infty^\text{n.c.} \rangle=  1.51\times \frac{N}{\nu}+ \frac{N(1-p)}{p\nu}, \ \ \ \langle \tau_1^\text{n.c.}\rangle  =    \frac{N}{p \nu}. \label{MFPTNC}
\end{equation}
Using these expressions to evaluate  (\ref{TvsQ})  and (\ref{ValueQ}), we obtain, in the regime $\Delta\ll \langle \tau_\infty^\text{n.c.}\rangle\sim N/(p\nu)$  and $p\ll1$,
\begin{equation}
\langle T^{\text{n.c.}}\rangle= \frac{N}{0.51 \ p^2 \nu}. \label{DiffRes}
\end{equation}
Comparing with the optimized value (\ref{OptT}), we see that $T_m /\langle T^{\text{n.c.}}\rangle\sim p\ll1$. 
This shows that surrounding the target by a crowded space, at the cost of considerably slowing the motion in the compartment, can actually lead to faster kinetics of multisite reactions. 
The fact that  $\langle T\rangle<\langle T^{\text{n.c.}}\rangle$ in some regimes can be checked in Fig.~\ref{Fig2RW}(c). Next, in Fig.~\ref{Fig2RW}(e), we show the acceleration factor $\langle T\rangle/\langle T^{\text{n.c.}}\rangle$,  whose minimal value scales as $p$, as predicted by the theory. Acceleration factors of the order of $\langle T\rangle/\langle T^{\text{n.c.}}\rangle\simeq 0.1$ are clearly seen. 
The acceleration effect is not limited to the regime $\Delta\gg \tau_c$ but also appears for smaller $\Delta$ as soon as $\Delta\gg\nu^{-1}$ [compare Eqs (\ref{TSmallerDelta}) and (\ref{DiffRes})]. 

 The acceleration factor is shown for various obstacle densities in Fig.~\ref{Fig2RW}(f). For $q>q_c$, the correlation length in the percolation cluster is finite, and the random walk becomes diffusive above this length scale. One may thus expect that $\langle T\rangle$ does not depend on $R$ when $q$ is very different from $q_c$, as seen in  Fig.~\ref{Fig2RW}(f). Nevertheless, for a finite range of $q$ and  small enough reactivity, the presence of the compartment accelerates  multisite reactions. 

\section{Generalizations}

\subsection{Multiple searchers and targets}
The theory can be generalized to the case of $n_s$  non-interacting searchers, initially with uniform distribution in the volume, that can all activate the target. In this case, a dual-site reaction can either result from two successive activation events by the \textit{same} random walker (mechanism $\text{I}$), or involve two distinct random walkers (mechanism $\text{II}$). Let us define $\lambda_i\ \delta t $ as the probability that an alive target reacts during $\delta t$ via the mechanism $i\in\{\text{I},\text{II}\}$, at a coarse grained level ($\delta t\gg \Delta$). Let us consider the case $\Delta\ll \langle \tau_\infty\rangle$. 
Clearly, $\lambda_\text{I} = n_s/\langle T\rangle$. For the mechanism involving two distinct searchers, $ \lambda_\text{II}= n_s(n_s-1)\Delta/\langle\tau_\infty\rangle^2$ is the product of the probability per unit time of an activation by one of the $n_s$ searchers ($n_s/\langle\tau_\infty\rangle$), and the probability of a second activation during $\Delta$ by the remaining $n_s-1$ searchers  $[(n_s-1)\Delta/\langle\tau_\infty\rangle]$. Using Eq.~(\ref{MeanTComp}), we show that $\lambda_\text{I}>\lambda_\text{II}$  is obtained when $n_s<n_s^*\sim \langle\tau_\infty^*\rangle/\Delta$. Interestingly, $n_s^*$ can be large compared to one when $\Delta$ is much smaller than the first passage to the target $\langle\tau_\infty^*\rangle$, meaning that there is a range of number (and thus, concentration) of searchers for which the mechanism $\text{I}$ is faster than the mechanism in which each reaction involves distinct searchers. This result also holds for the reaction kinetics of many targets, possibly in distinct compartments, since the searchers spend most of their time outside the compartments.   

\subsection{Multisite reactions with more than two sites}
Next, we can also generalize the theory to the case where the target has $m$ sites that need to be all activated during a time window $\Delta$, with $m\ge2$. For simplicity we focus on the case $\Delta\gg \tau_c$. We argue that the mean reaction time in this regime is given 
 by (\ref{TvsQ}), where $Q$ is the probability of activating $m-1$ times the target without escaping to large distances, so that $Q=(1-\langle\tau_1\rangle/\langle\tau_\infty\rangle)^{m-1}$. This leads to 
\begin{align}
\langle T\rangle=N \frac{\left(\alpha R^{d_\text{w}-d_\text{f}}+\frac{1}{p\nu}\right)^m}{(\alpha R^{d_\text{w}-d_\text{f}})^{m-1}}.
\end{align}
Optimizing with respect to $R$ leads to the optimal reaction time
\begin{align}
\langle T\rangle_m=N \frac{m^m}{(m-1)^{m-1}\nu p} \sim \frac{N}{p\nu}
\end{align}
In turn, in the case without compartment, if $\Delta\ll N/(p\nu)$, still with $p\ll 1$, we have 
\begin{align}
\langle T\rangle_{\text{n.c.}}=\frac{\langle \tau_\infty\rangle^m}{(\langle \tau_\infty\rangle-\langle \tau_1\rangle)^{m-1}}=\frac{N}{\nu (0.51)^{m-1} p^m }.
\end{align}
 Comparing the last  two expressions, we obtain a speed-up factor $\langle T\rangle/\langle T\rangle_{\text{n.c}}\sim p^{m-1}$, which  drastically decreases with $m$ in the weakly reactive case ($p\ll 1$). 
 
  \begin{figure}[ht!]
    \includegraphics[width=0.9\linewidth]{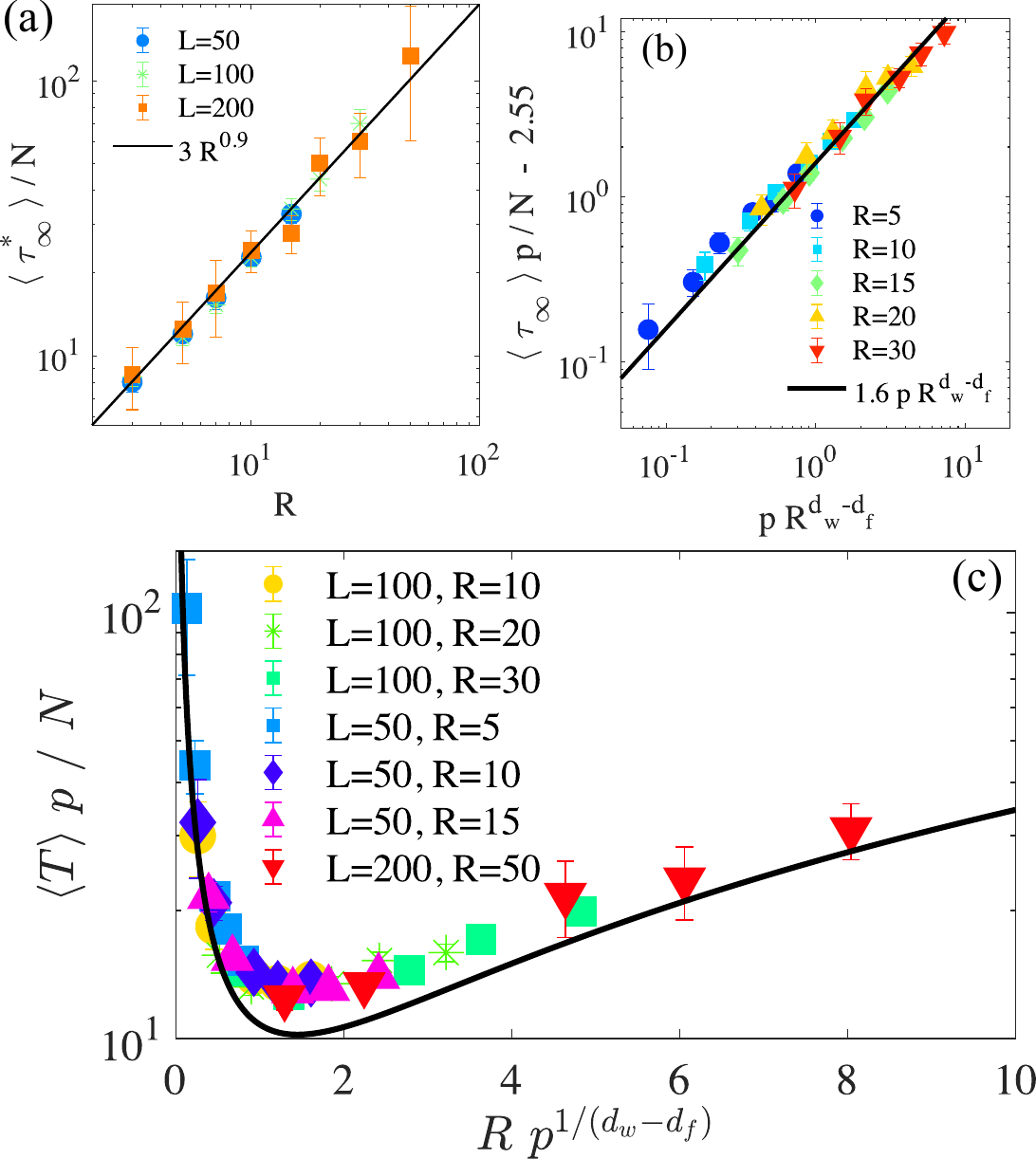}
    \caption{Results for mobile targets, with $q=0.311$. (a) Mean first passage time versus compartment radius $R$. (b) Validation of the expression (\ref{ExprATSmallReactMT}) for the activation time for weak reactivity ($p<0.1$). (c) Reaction time versus $R$ for different reactivities  and $\Delta=10^4$ for $L=50$, $\Delta=10^5$ for $L=100$, $\Delta=10^6$ for $L=200$  and reactivities $p<0.1$. The line represents the prediction (\ref{ExprReactionTSmallReactMT}).}
    \label{fig:MobileTarget}
\end{figure}

\subsection{Mobile targets}
\label{Mobile}

We  generalize  our results to the case of mobile targets: after each step performed by the searcher, the target performs one move before the next move of the searcher, with the constraint that it remains in the compartment. Encounters can happen everywhere within the compartment. 
For the critical obstacle density $q\simeq 0.311$,  the mean FPT behaves as $\langle \tau_\infty^*\rangle \sim N R^{\gamma}$ with $\gamma\simeq 0.9 $ [Fig. \ref{fig:MobileTarget}(a)], which is smaller than $d_\text{f}-d_\text{w}$ for immobile targets. This is probably due to the fact that the probability of encounter is not uniform among all sites in the compartment, but may more likely occur near the compartment boundary. However, in the limit of small reactivity, we find 
\begin{align}
\langle \tau_\infty\rangle \simeq 2.55\frac{N}{p} + 1.6 \ N R^{d_\text{f}-d_\text{w}} \label{ExprATSmallReactMT},
\end{align}
see Fig. \ref{fig:MobileTarget}(b). Here,  the form of the first term is expected from Kac's theorem, while the correction can be understood by noting that, at leading order for small $p$, the distribution of positions at the encounter is uniform inside the compartment. Hence, after an encounter the distance between target and searcher displays the same behavior as for an immobile target for times of order $\tau_c$, so that one can expect that the correction to the $2.55\frac{N}{p}$ term satisfies the same scaling as for immobile targets. Next, $\langle \tau_1\rangle$ is given by (\ref{ExprATSmallReactMT}) for $R=0$. Now, if one focuses on the  weakly reactive limit ($p\le 0.1$), and one applies Eq. (\ref{MeanTComp}), one obtains 
\begin{equation}
\langle T \rangle  \simeq  N \frac{\left(1.6\ R^{d_{\text{w}}-d_{\text{f}}}+\frac{2.55}{p}\right)^2}{ 1.6 R^{d_{\text{w}}-d_{\text{f}}}}.  \label{ExprReactionTSmallReactMT}
\end{equation}
This expression is in good agreement with simulations, see Fig. \ref{fig:MobileTarget}(c). The reaction time displays the same scaling as in the case of immobile target, so that the acceleration effect due to the compartment also holds for mobile targets.

\subsection{Presence of passive states} 
The last generalization is the case of targets that are mobile within the compartment, and to the case where the random walker becomes passive after an activation event during a certain lag-time    that is exponentially distributed with mean $\delta_0$. We find in Appendix \ref{SecPassive} that acceleration can still be obtained if this lag-time is much smaller than $\tau_c$: in this case, if $\Delta\gg \tau_c$, then Eq. (\ref{MeanTComp}) is still valid, while in the opposite case where $\Delta\ll \tau_c$, one obtains 
 \begin{equation}    
\expval{T} =  f_\theta(\delta_0 /\Delta) \frac{b N}{p^2 \Delta^\theta},
\end{equation}
where $b$ is the same prefactor as in Eq.~(\ref{TSmallerDelta}), and the finite function $f_\theta$ is given in Eq (\ref{fTheta}). Hence, in the regime $\delta_0,\Delta\ll \tau_c$, taking into account a passive state of the random walker where the activation is impossible only modifies the reaction time by a numerical prefactor. 

\section{Concluding remarks} 

 Our  results extend to a compartment geometry, and at the level of search times of single molecules, the studies on the effect of transport on multisite reactions in Refs.~\cite{takahashi2010spatio,gopich2013diffusion,hellmann2012enhancing}. In the case of diffusive transport inside the compartment, the mean reaction time is given by (\ref{DiffRes}), independently of the compartment size which cannot be optimized. In this case, a non-monotonic variation of $\langle T\rangle$ with the diffusion constant can be obtained if there is a fixed local activation rate $k$ on the target [with $p=k/(k+\nu)$], in agreement with Refs.~\cite{takahashi2010spatio,gopich2013diffusion} (see Appendix \ref{SinkReact} for details). The acceleration effect described in our work occurs at fixed $p$ and is therefore more general.

We have studied multisite reactions, which occur when a target  is successfully hit several times within a restricted time window, in the context of compartmentalized media. Considering  compartments with fractal-like subdiffusion, we have shown that, in spite of the crowding induced slow-down of motion, an acceleration of multisite reactions can be obtained,  as a result of dynamic effects, without an energetic gain for the random walker to be inside the compartment. This phenomenon happens in the large volume limit (a realistic hypothesis if compartments are small compared to the surrounding confining volume) and 
in the weakly reactive regime. This is a generic situation ~\cite{bar2011moderately,grebenkov2019imperfect,grebenkov2023diffusion} that occurs, for example,  when reactive sites occupy a small surface of the target molecule and may be relevant for phosphorylation~\cite{takahashi2010spatio}. 
We show in Appendix \ref{GenViscoElasticSub} how these results can be generalized to viscoelastic subdiffusion, as a consequence of the similar scalings of first passage quantities for fractal and viscoelastic subdiffusion~\cite{guerin2016mean,Condamin2007,levernier2019survival,Krug1997,Molchan1999,meroz2011distribution,levernier2018universal,mendes2024imperfect} (except for strongly non-stationary initial states~\cite{ReviewBray,Levernier2022Everlasting}). 
It would be interesting to determine whether similar acceleration effects could occur in more complex models of transport in disordered media, including binding energies to obstacles. 
Although our theory uses simplified models of transport in crowded media, which differ from real complex situations such as MLOs, our study reveals, as a general principle, that a loss of mobility in crowded compartments may not always be seen as a difficulty to be overcome (by over-concentrating reactants), but can actually speed up complex reactions such as multi-site reactions, especially in the weakly reactive regime.

\begin{acknowledgments}
We acknowledge the support of the grant \textit{ComplexEncounters}, ANR-21-CE30-0020. We thank Olivier Bénichou for interesting discussions. Computer time for this study was provided by the computing facilities MCIA (Mesocentre de Calcul Intensif Aquitain) of the Universit\'e de Bordeaux and of the  Universit\'e de Pau et des Pays de l’Adour. 
\end{acknowledgments}
 
The data that support the findings of this article are openly available \cite{DataComp}. 
  
\appendix

\section{Remarks on Euclidian and chemical distances}
\label{Dist}
To study random walks in fractals it is common to use the ``chemical'' (or topological) distance, defined as the length of the shortest path between two points, rather than the Euclidian distance. Indeed, the critical percolation cluster with Euclidian distance displays multifractal properties \cite{bunde1990multifractal}. However, we argue that Eq.~(\ref{ScalingMFPTInitialDistance}) for the mean first passage time to the central site is still valid. Let us denote by $\ell$ the chemical distance. In chemical distance space, we have 
\begin{equation}
P(\ve[0], t\vert \ve[r]) \simeq \frac{1}{t^{d_{\text{f}}'/d_{\text{w}}'}}\Pi_0\left(\frac{\ell}{t^{1/d_{\text{w}}'}}\right), \label{ScalingFractalChem}
\end{equation} 
where $d_{\text{f}}'$ and $d_{\text{w}}'$ are respectively the spatial and   walk dimension for the chemical distance. Using chemical distances, this propagator is a self-averaging quantity. This enables us to derive the mean first passage time when the starting position is far from the compartment boundary, with a fixed chemical distance to the target site $\ell$:
\begin{align}
\frac{\langle \tau_{\ell}^* \rangle }{N}  &=  \int_0^\infty dt\  [P(\ve[0],t\vert \ve[0])-P(\ve[0],t\vert \ve[r])]\nonumber\\
&\simeq A_0 \ \ell^{d_{\text{w}}'-d_{\text{f}}'}\label{TauEllStar},
\end{align}
with $A_0=\int_0^\infty u^{-\frac{d_{\text{f}}'}{d_{\text{w}}'}}[\Pi_0(0)-\Pi_0(u^{-1/d_{\text{w}}'}))]$. Next, the probability distribution function of Euclidian distances between two points, given that the chemical distance is $\ell$, reads \cite{neumann1988distributions,porto1998probability}
\begin{align}
\Phi(r\vert \ell) =\frac{1}{r}f(r/\ell^\mu) \label{ScalingPhi},
\end{align}
where $f$ is a scaling function and $\mu=d_{\text{w}}'/d_{\text{w}}=d_{\text{f}}'/d_{\text{f}}$. Note that here we use the normalization $ \int_0^\infty dr \Phi(r\vert \ell)=1$. We can write a relation between the mean first passage time for fixed $\ell$ and the mean first passage time for fixed $r$:
\begin{align}
\langle \tau_{\ell}^* \rangle=\int_0^\infty dr \  \Phi(r\vert \ell) \ \langle \tau_{r}^* \rangle\label{IntEq}.
\end{align} 
This is an integral equation for $\langle \tau_{r}^* \rangle$. Since $\langle \tau_{\ell}^* \rangle$ is given by Eq.~(\ref{TauEllStar}), and $\Phi$ satisfies the scaling (\ref{ScalingPhi}), the above integral equation admits the obvious solution
 \begin{align}
\frac{\langle \tau_{r}^* \rangle }{N}  & \simeq A \ r^{d_{\text{w}}-d_{\text{f}}}\label{TauRStar},
\end{align}
where 
\begin{align}
A=\frac{A_0}{\int_0^\infty d\rho f(\rho) \rho^{d_{\text{w}}-d_{\text{f}}-1}}.
\end{align}
The solution  (\ref{TauRStar}) is an acceptable solution of Eq.~(\ref{IntEq}) only if $A$ exists. It is known that $f$ decays as a stretched exponential at large arguments, whereas for small $\rho$ one has $f(\rho)=\rho^{g_3}$ with $g_3\simeq 3.98$ for critical percolation  in $d=3$ \cite{neumann1988distributions,porto1998probability}, so that $f(\rho) \rho^{d_{\text{w}}-d_{\text{f}}-1}\sim \rho^{4.24}$ which is integrable at small $\rho$. We conclude that $A$ exists and that the law (\ref{TauRStar}) holds for the Euclidian distance space, in spite of the multifractal character of the critical percolation cluster in this case. Note that this law is expected to hold when $1\ll r \ll R$. For chemical distance space, formulas of the form (\ref{TauEllStar}) often hold up to $\ell=1$ \cite{benichou2008zero,guerin2021universal}, a property which is not satisfied with Euclidian distances, however this can be understood since for $r=1$ chemical and Euclidian distances coincide, whereas, for larger $r$, the chemical distance to the target  can be much larger than $r$. 

\section{Calculation details}
\label{CalculationDetails}
\subsection{Reaction time when $\Delta$ is comparable to $\langle\tau_\infty\rangle$}
\label{Deriv_Eqs}

Here, we focus on the regime where $\Delta$ is large compared to the time to leave the compartment ($\Delta\gg \tau_c\equiv(R/\sqrt{K})^{d_w/2}$), but possibly on the order of $\langle\tau_\infty\rangle$. Since $\Delta\gg\tau_c  $, if a second activation has not occurred at a time $\Delta$ after a first activation, the search for a reaction even has to start over entirely. Hence, the time for a dual-site reaction can be written as the resetting problem:
\begin{equation}
T=\tau_\infty+ 
\begin{cases}
 \tau_1 & (\text{if } \tau_1<\Delta),\\
\Delta+ T' & (\text{if } \tau_1>\Delta),
\end{cases}
\end{equation} 
where $T$ and $T'$ are independent realizations of the dual-site reaction time. 
 The above equation means that  $T$ belongs to the widely studied class of search processes with resetting \cite{evans2020stochastic,reuveni2014role}. In particular, it has appeared in Refs~\cite{pal2017first,reuveni2016optimal}. 
 Following \cite{pal2017first}, we write
\begin{equation}
    T = \tau_\infty + \min(\tau_1,\Delta) + I\{\Delta\leq \tau_1\} T',
\end{equation}
where the indicator function $I\{\Delta\leq \tau_1\}$ is equal to unity if $\Delta \leq \tau_1$ and zero otherwise. Averaging the above equation and using  the fact that $T'$ and $\tau_1$ are statistically independent, one obtains 
\begin{equation}
    \langle T \rangle = \frac{\langle \tau_\infty \rangle + \langle\min(\tau_1,\Delta)\rangle}{1- \text{Prob}(\tau_1\ge \Delta)}.\label{94325}
\end{equation}
We note that $\text{Prob}(\tau_1\ge t)=S_1(t)$ is the survival probability of a random walker that is started close to the target. 
Since $\Delta \gg \tau_c$, only the trajectories that leave the compartment (which happens with probability $\pi_1$) are relevant for the calculation of $S_1(\Delta)$. Since the movement of the random walker outside the compartment is diffusive in three dimensions, the survival probability for large $t$ can simply be written as
\begin{equation}
    S_1(t)  \simeq \pi_1\, e^{-t/\langle \tau_\infty \rangle}\quad \text{ for }\quad t\gg \tau_c. \label{S1}
\end{equation}
The above expression omits the weight of the direct trajectories, that reach successfully the target before escaping to large distances outside the compartment, which last much shorter times than $\langle \tau_\infty\rangle$ in the large volume limit. Hence, the average activation time $\langle\tau_1\rangle=\int_0^\infty dt S_1(t)$ can be evaluated using the above equation~\cite{Benichou2010}, so that 
\begin{align}
\pi_1=\langle\tau_1\rangle/\langle\tau_\infty\rangle.\label{Pi1}
\end{align}
Next, the quantity $\langle \min(\tau_1,\Delta) \rangle$ can be calculated as
\begin{align}
    \langle \min(\tau_1,\Delta) \rangle = \int_0^\Delta dt \ F_1(t) t+\int_\Delta^\infty dt \ F_1(t)\Delta,
    \end{align}
where $F_1(t)=-\partial_t  S_1(t)$ is the probability distribution function (PDF) of $\tau_1$. Integrating by parts   the first integral of the above equation, and using Eqs.~(\ref{S1}) and (\ref{Pi1}), we obtain
\begin{align}
    \langle \min(\tau_1,\Delta) \rangle= \int_0^\Delta dt \ S_1(t)= \langle\tau_1\rangle (1-e^{-\Delta/\langle \tau_\infty\rangle}).  \label{mintau1Delta}
\end{align}
Using the  expressions (\ref{S1}), (\ref{Pi1}) and (\ref{mintau1Delta}), we can write (\ref{94325}) as
\begin{equation}
    \langle T\rangle=\frac{\langle\tau_\infty\rangle+\langle\tau_1\rangle\left(1-e^{-\Delta/\langle\tau_\infty\rangle}\right)}{1-\frac{\langle\tau_1\rangle}{\langle\tau_\infty\rangle} e^{-\Delta/\langle\tau_\infty\rangle}}. \label{TNotSmallDelta}
\end{equation}
As expected, when $\Delta\gg\langle\tau_\infty\rangle$, the above expression reduces to $\langle T\rangle=\langle \tau_\infty\rangle +\langle \tau_1\rangle$. In the opposite limit $\Delta\ll \langle \tau_\infty\rangle$, one recovers Eq.~(\ref{MeanTComp}).

\subsection{Derivation of the reaction time for small values of $\Delta$ [Eq (\ref{TSmallerDelta})] }

\label{DerivSmallDelta}

We consider here the reaction time when $\Delta\ll \tau_c$. We first calculate $Q$, previously defined as the probability that a reaction will be completed before the random walker escapes far from the compartment, given that the first activation has just occurred. Two scenarios can distinctly be identified: 
(i) The second activation occurs within a time window $\Delta$ after the first one, which corresponds to an event with probability $Q_1=1-S_1(\Delta)$,
(ii) $n$ activations happen after the first one without an escape to infinity but none of them happens with a time interval smaller than $\Delta$, which corresponds to a probability $B_n = [(1-P_\infty) (1-Q_1)]^n$ and then, the $(n+1)^{th}$ activation occurs within the $\Delta$ window, which happens with probability $Q_1$.  Here, $ P_\infty$ is the probability, starting at a time $\Delta$ after an activation, to escape to large distances before activating the target. Therefore, the probability $Q$ that a reaction is completed before the walker escapes to infinity can be written as $Q=Q_1 + Q_1 \sum_{n=1}^\infty B_n $, so that
\begin{equation}
    Q = \frac{Q_1}{1-(1-P_\infty)(1-Q_1)} = \frac{1-S_1(\Delta)}{1 - (1-P_\infty) S_1(\Delta)}.\label{Qred}
\end{equation}

Now, when $\Delta\ll \tau_c$, the typical distance $r$ traveled during $\Delta$ is small compared to $R$, so that the activation time, starting at a time $\Delta$ after having touched the target, is well approximated by $\langle\tau_1\rangle$. Calculating $P_\infty$ as the ratio of activation times leads to
\begin{align}
P_\infty(\Delta)=  \frac{\langle\tau_1\rangle}{\langle\tau_\infty\rangle}. \label{PiDelta} 
\end{align}

Let us now calculate $S_1(\Delta)$. We introduce $k$ such that $p=k/(k+\nu)$, as the probability per unit time to activate the target when the random walker is on the target site. For small reactivity, $k \simeq p \nu$. One can write $F_1(t)=k P_a(\ve[r]_T,t\vert\ve[r]_1)$, where   $P_a(\ve[r]_T,t\vert \ve[r]_1)$ is the probability to be at the target without any   activation during  $]0;t[$, starting at a neighboring site $\ve[r]_1$ of the target at time $0$. We argue that, if $1-S_1(t)$ is small enough, which occurs for small enough $t$ and small enough $k$, we may replace $P_a(\ve[r]_T,t\vert \ve[r]_1)$ by $P(\ve[r]_T,t\vert \ve[r]_1)$ which is the propagator in the absence of absorption at the target. With this approximation, we obtain
\begin{align}
F_1(t)\simeq k \ P(\ve[r]_T,t\vert \ve[r]_1)\simeq  \frac{p\ \nu\ \Pi (0)}{t^{d_\text{f}/d_\text{w}}},
\end{align}
where the second equality follows from Eq.~(\ref{ScalingFractal}) when $t\gg \nu^{-1}$. The above expression generalizes a similar scaling behavior obtained in Ref.~\cite{mercado2019first} for diffusive random walks in one-dimension for imperfect targets ($d_\text{f}=1,d_\text{w}=2$). Hence, the following approximation
\begin{align}
1-S_1(\Delta)\simeq \left(1-\frac{d_\text{f}}{d_\text{w}}\right)\  \Pi (0) \ p\ \nu \Delta^{1-d_\text{f}/d_\text{w}}  \label{1MoinsS}
\end{align}
is valid under the self-consistent condition that $1-S_1(\Delta)\ll1$. 
Collecting the expressions (\ref{1MoinsS}), (\ref{PiDelta}),  (\ref{Qred}) and (\ref{MFPTImp}), we evaluate $\langle T\rangle$ in Eq.~(\ref{TvsQ}) in the regime of small $\Delta$, with $p\ll 1$ to be
\begin{align}
\langle T\rangle\simeq 
 \frac{\langle\tau_\infty\rangle P_\infty}{1-S_1(\Delta)}\simeq \frac{b\ N }{p^2  \Delta^{1-d_\text{f}/d_\text{w} }}, \label{Eq26}
\end{align}
where $b$ is a prefactor that does not depend on $N,R,p$.

\section{Details on simulations}
\label{Simulations}
\textbf{Details on the simulated system.}
In our simulations, we consider a cubic lattice with $L^3$ sites, occupying a three-dimensional volume of size $L\times L \times L$. A target site is chosen at $(L/2,L/2,L/2)$. For all sites located at less than the Euclidian distance $R$ from the central site, we determine if the site is obstructed (with probability $1-q$) or free (with probability $q$). Next, we identify all sites that  are connected to the medium outside the compartment which are, in our simulations, the sites that are accessible to the random walk. This enables us to avoid starting a random walk in a site that is not connected to the exterior. We assume that the target is accessible from the exterior medium, hence if the target site is not connected to it, we generate a new configuration of the compartment. 
By convention, to represent the confining boundaries, we add one layer of obstructed sites at the external boundaries of the domain, so that all free sites have  six neighbors, obstructed or not. Random walks are considered on the obtained network: at each time step the random walker attempts to jump to one of its neighboring sites at random. If the new site is free, the jump is accepted, otherwise the jump is rejected and another attempt is made at the next time step. This corresponds to the ``blind ant algorithm''. 
The time step is taken as the unit of time, the distance between two neighboring sites is the unit of length. 

\textbf{Estimate of error-bars.}
All measured average quantities in simulations are averaged over both ensemble and disorder. More precisely, consider the case where we want to measure the average of a random variable $X$ (for example, a first passage time), given that we have access to $n_r$ configurations of the compartment ($n_r$ realizations of the disorder), and that for each realization $j\in\{1,...,n_r\}$ one has access to $n_j$ measures of $X$, $X_i^{(j)}$ with $1\le i\le n_j$. The value to estimate is 
$m=\overline{\langle X\rangle}$, where the brackets $\langle \cdot\cdot\cdot\rangle$ denote the ensemble average and over-lines $\overline{\cdot\cdot\cdot}$ represent the average over the disorder. As an estimator of $m$ we use 
\begin{align}
&X_e = \frac{1}{N_e}\sum_{j=1}^{n_r}  \sum_{i=1}^{n_j}X_i^{(j)},
\end{align}
where $N_e=\sum_{j=1}^{n_r} n_j$ is the total number of recorded events. Obviously, $\overline{\langle X_e\rangle}=m$, and to calculate the variance we write 
\begin{align}
&X_e-m = \frac{1}{N_e}\sum_{j=1}^{n_r}  n_j (Y_j-m),
\end{align}
where $Y_j=n_j^{-1}\sum_{i=1}^{n_j}X_i^{(j)}$ is the empirical average of $X_i^{(j)}$ for the configuration $j$. 
Squaring $(X_e-m)$, taking an ensemble average  and using the fact that the variables $Y_k$ are statistically independent leads to
\begin{align}
\langle (X_e-m)^2\rangle = \frac{1}{N_e^2}\Bigg(&\sum_{j,k=1}^{n_r}  n_j n_k (\langle Y_j\rangle-m)(\langle Y_k \rangle-m) \nonumber\\
&+\sum_{j=1}^{n_r}  n_j^2 [ \langle Y_j^2\rangle-\langle Y_j\rangle^2] \Bigg) .
\end{align}
Note that if $\sigma_j=\text{var}(X_i^{(j)})$ is the variance of $X$ for the configuration $j$, then $\langle Y_j^2\rangle-\langle Y_j\rangle^2=\sigma_j/n_j$. We  used the above formula to estimate the $95\%$ confidence intervals due to statistical uncertainty for a given number of configurations, replacing $\langle Y_i\rangle$ by the empirical average of the measures of $X$ for given configuration, and $\sigma_i$ their variance. In the case that all $n_j=n_e$ are equal, we obtain
\begin{align}
&\overline{\langle (X_e-m)^2\rangle}=   \frac{1}{n_r} \overline{(\langle Y_j\rangle-m)^2}  +   \frac{1}{N_e} \overline{\sigma}  
\end{align}
This formula indicates how the difference between $X_e$ and $m$ depends on the average variance of $X$ for a given realization and also the dispersion of the averages $\langle Y_i\rangle$ obtained for each realization.

\begin{figure}[ht!]
    \centering
    \includegraphics[width=\linewidth]{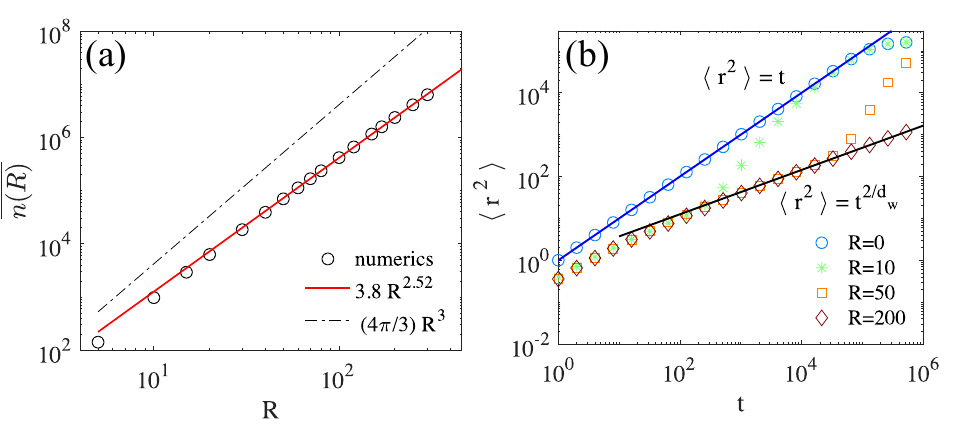}
    \caption{(a) Symbols: measured average number of sites of the compartment $\overline{n(R)}$ that are connected to the external medium. We also show the power-law behavior $n(R)\sim R^{d_{\text{f}}}$ with $d_{\text{f}}\simeq 2.52$ (red line) and the volume occupied by the compartment $4\pi R^3/3$ (dash-dotted line). 
(b) Symbols: mean square displacement for a random walker starting at the center of the compartment ($r=0$), with $L=800$ and $R$ indicated in the legend. We also show the power-law $\langle r^2\rangle=t^{2/d_{\text{w}}}$ with $d_{\text{w}}\simeq 3.78$ (black line) and the diffusive law $\langle r^2\rangle =t$ (blue line).}
    \label{fig:checkDim}
\end{figure}

\textbf{Check of the spatial and   walk dimensions $d_\text{f}$ and $d_\text{w}$.}
The spatial and walk dimensions are usually considered for the largest clusters of a percolation network, where $d_{\text{f}}\simeq 2.523$ \cite{jan1998random,ballesteros1999scaling,deng2005monte,benAvraham2000} and $d_{\text{w}}\simeq3.78$ 
\cite{argyrakis1984random,blavatska2008walking,ben1982diffusion,havlin1983diffusion}. In  Fig.~\ref{fig:checkDim}, we checked numerically that the same values of $d_\text{f}$ and $d_\text{w}$ also describe  the spatial and walk dimensions in our system, where the random walk occurs on sites connected to the exterior of the compartment.

\section{Random Walker that can switch into a passive state}
\label{SecPassive}

Here, we assume that the random walker undergoes some internal alteration at each activation that prevents it from activating the target again during a lag-time $\delta$, after which the random walker returns to its regular searching state. We assume that $\delta$ is exponentially distributed with average $\delta_0$. If $\delta_0$ is large compared to the time to leave the compartment $\tau_{c}$, consecutive activations are not possible and no acceleration effect can occur. We thus focus on the case  $\delta_0\ll \tau_c$, and we investigate the regimes $\Delta \ll \tau_c$ and $\Delta \gg \tau_c$.

In the case   $ \tau_c\ll\Delta\ll\langle  \tau_\infty\rangle$, one has $\langle T\rangle=\langle\tau_\infty\rangle/Q$ where $Q$ is the probability to activate the target before escaping to large distances, starting from the neighborhood of the target. Since $\delta_0 \ll \tau_c$, the probability that an activation time is smaller than $\delta$ is low and affects only moderately  $Q$, so that Eq.~(\ref{MeanTComp}) remains valid. Let us now analyze the situation where $\Delta \ll \tau_c$. Following the method described in Appendix \ref{CalculationDetails}, we first calculate $Q_1$, defined as the probability that, after an activation, a second activation happens before the time $\Delta$. Defining $t_i$ as the $i$-th realization of the activation time starting from the neighborhood of the target, we write
\begin{align}
    Q_1& = \text{Pr}(\delta < t_1 < \Delta) + \text{Pr}(t_1 < \delta; \delta < t_1 + t_2 < \Delta) \nonumber\\
    &+ \text{Pr}(t_1 + t_2 < \delta; \delta < t_1 + t_2 +t_3 < \Delta) + ... 
    \label{eq:A_Condition}
\end{align}
In principle, one would have to calculate $Q_1$ as an infinite sum. However, when reactivity is low and $\Delta\ll \tau_c$ the probability $\text{Pr}(t_1 < \Delta)$ is small, and the probabilities of the next terms in the above series are even smaller, so that one can approximate $Q_1\simeq \text{Pr}(\delta < t_1 < \Delta)$ by the first term of the series. Therefore, we can write
\begin{equation}
    Q_1 \simeq \int_0^{\infty} \frac{e^{-\delta/\delta_0}}{\delta_0}  \Theta(\Delta-\delta) \qty[S_1(\delta)-S_1(\Delta)] d\delta,
\end{equation}
where $\Theta(\cdot)$ represents the Heaviside step function. Using the expression of $S_1(t)$ [Eq.~(\ref{1MoinsS})], $Q_1$ can be rewritten as
\begin{equation}
    Q_1 = \frac{ p\nu \theta\  \Pi(0) \Delta^{\theta} }{ f_\theta(\delta_0/\Delta) },
\end{equation}
where $\theta=1-d_\text{f}/d_\text{w}$, and
\begin{equation}
    f_\theta(x) = \frac{1}{1-e^{-1/x} - x^\theta \int_0^{1/x} y^{\theta}e^{-y} dy}, \label{fTheta}
\end{equation}
which is a positive function of $x$ with $f(0)=1$. Then, since the mean multisite reaction time $\expval{T}$ is simply $\expval{T} = \expval{\tau_\infty}\frac{1-(1-Q_1)(1-P_\infty)}{Q_1}$, with   $P_\infty = \expval{\tau_1}/\expval{\tau_\infty}$, one can write
\begin{equation}    
\expval{T} =  f_\theta(\delta_0/\Delta) \frac{b N}{p^2 \Delta^\theta},
\end{equation}
where $b$ is the same prefactor as in Eq.~(\ref{Eq26}). Hence, in the regime $\delta_0,\Delta\ll \tau_c$, taking into account a passive state of the random walker where the activation is impossible only modifies the reaction time by a numerical prefactor.

\section{Reaction time for sink reactivity and compartments with diffusive transport}
\label{SinkReact}
Here we briefly investigate the case where one uses an activation rate $k$ on the target site instead of assuming a fixed activation probability $p$ at each passage to the target. Here $k$ is defined so that an activation event occurs during $[t;t+dt]$ with probability $k \ dt $ when the random walker is on the target at $t$. Since the time spent on the target at each passage is exponentially distributed with average $\nu^{-1}$,  the probability to activate the target at each passage is
\begin{align}
p=\frac{k}{k+\nu}.
\end{align}
In the case where the transport inside the compartment is diffusive, we argue that the reaction time is given by Eq.~(15) irrespective of the size of the compartment (as soon as $R\gg1$). This is justified since $\langle\tau_r\rangle$ saturates for $r\gg1$ so that it is nearly constant as soon as $r\gg1$, including at the compartment's interface $r=R$. The reaction time in this case reads
\begin{equation}
\langle T\rangle= \frac{N}{0.51 \ p^2 \nu} =\frac{N}{0.51 \  \nu}\left( \frac{k+\nu}{k} \right)^2.
\end{equation}
At fixed $k$, $\langle T\rangle$ admits a minimum when seen as a function of $\nu$. Since the local diffusivity is proportional to $\nu$, this formula means that the reaction time is non-monotonic with the local diffusivity, which is consistent with the results of Refs.~\cite{takahashi2010spatio,gopich2013diffusion}. Note, however, that when $p$ is fixed, $\langle T\rangle $ can no longer be optimized by varying the diffusion constant (or the compartment radius), whereas, in the case of fractal subdiffusion in the compartment, $\langle T\rangle $ can be optimized by varying the compartment size. The optimization effect investigated in our paper is therefore more general than the one identified for diffusive search and can occur even for fixed low values of $p$.

\section{Generalization  to the case of viscoelastic subdiffusion in the compartment}
\label{GenViscoElasticSub}
 Here, we use scaling arguments to generalize the theory to the case of viscoelastic subdiffusion inside the compartment. We consider a continuous space of dimension $d=3$. We assume that the mean-square displacement inside the compartment scales as $\langle [x(t)-x(0)]^2\rangle\propto Kt^{2/d_w}$, whereas outside the compartment transport is diffusive, with diffusion coefficient $D$. The target is a sphere of radius $a$ (with $a\ll R$), at the surface of which activation occurs with surface reactivity $\kappa$. The compartment, of radius $R$, is embedded inside a confining domain of volume $V$. 

In Ref.~\cite{mendes2024imperfect}, it was found that, if $d_w<d+2$ then, for small (sink) reactivity, the activation time can be written 
as the sum of a reaction controlled time and a time displaying the same scaling as the mean FPT. Extrapolating that this property also holds for surface reactivity, 
we estimate that, for $r\ll R$, in the small reactivity limit,
\begin{align}
\langle\tau_r\rangle\simeq\frac{V}{4 \pi a^2 \kappa}+\frac{A V (r^{d_w-3}-a^{d_w-3})}{K^{d_w/2}},
\end{align}
with $A$ a numerical prefactor. Here we have assumed that the stationary probability density is uniform $p_s(\ve[x])=1/V$, as would be the case for viscoelastic subdiffusion in the absence of energy gain to be in the compartment. For larger $r>R$, invoking the continuity of the mean activation time at $r=R$, and assuming that the above scaling holds up to $r=R$, we have
\begin{align}
\langle\tau_r\rangle\simeq\frac{V}{4\pi a^2 \kappa}+\frac{\alpha V R^{d_w-3}}{K^{d_w/2}}+\frac{V}{4\pi D }\left(\frac{1}{R}-\frac{1}{r}\right),
\end{align}
with $\alpha$ a numerical coefficient. 
We assume that transport inside the compartment is slower than outside, including at the scale of the target. In this case, the last term is negligible, leading to
\begin{align}
\langle\tau_\infty\rangle\simeq\frac{V}{4\pi a^2\kappa}+\frac{\alpha V R^{d_w-3}}{K^{d_w/2}}.
\end{align}
Next, the mean activation time $\langle\tau_1\rangle$ when the initial position is the target surface reads
\begin{align}
\langle\tau_1\rangle\simeq\frac{V}{4\pi a^2\kappa}.
\end{align}
Applying the same arguments as in the main text to calculate the dual-site reaction time, we obtain in the regime $(R^2/K)^{d_w/2}\ll \Delta \ll \langle \tau_\infty \rangle$
\begin{align}
\langle T \rangle\simeq\frac{V\left(\frac{1}{4\pi a^2\kappa}+\frac{\alpha  R^{d_w-3}}{K^{d_w/2}}\right)^2}{\frac{\alpha  R^{d_w-3}}{K^{d_w/2}}}.
\end{align}
As in the case of fractal subdiffusion, there is an optimal  value of $R=R_m$ for which the mean reaction time is minimal:
\begin{align}
&R_m\sim \left(\frac{K^{2/d_w}}{a^2 \kappa}\right)^{\frac{1}{d_w-3}}, &\langle T\rangle_{R=R_m}\sim \frac{V}{a^2\kappa}.
\end{align}
For the same problem without compartment (nc), we have
\begin{align}
&\langle \tau_1\rangle^{nc}=\frac{V}{4\pi a^2 \kappa}, 
&\langle \tau_\infty\rangle^{nc}=\frac{V}{4\pi a^2 \kappa}+\frac{V}{4\pi D a}
\end{align}
so that
\begin{align}
\langle T \rangle^{nc}\simeq\frac{V\left(\frac{1}{4\pi a^2 \kappa}+\frac{1}{4\pi D a}\right)^2}{\frac{1}{4\pi D a}}
\end{align}
For low reactivity $\kappa\to0$, we thus have
\begin{align}
\langle T \rangle^{nc}\simeq V\frac{D}{4\pi a^3 \kappa^2}
\end{align}
We thus find that $\langle T \rangle^{nc}\sim \kappa^{-2}$ whereas with compartments the optimal time scales as $\langle T\rangle \sim \kappa^{-1}$. These scaling arguments suggest that the acceleration effect identified in this paper is still present in the case of viscoelastic subdiffusion.

 %

\end{document}